\documentclass[sigconf]{acmart}

\AtBeginDocument{%
  }

\setcopyright{acmlicensed}
\copyrightyear{2018}
\acmYear{2018}
\acmDOI{XXXXXXX.XXXXXXX}

\acmConference[Conference acronym 'XX]{Make sure to enter the correct
  conference title from your rights confirmation email}{June 03--05,
  2018}{Woodstock, NY}

\acmISBN{978-1-4503-XXXX-X/2018/06}

\usepackage{algorithm}
\usepackage{algpseudocode}
\usepackage{hyperref}
\usepackage[utf8]{inputenc}
\usepackage{tcolorbox}
\usepackage{lipsum}
\usepackage{multirow}
\usepackage{tabularx}
\usepackage{booktabs}  
\usepackage{graphicx}
\usepackage{subcaption}

\AtEndPreamble{
	\usepackage{hyperref}
	\hypersetup{
		colorlinks = true,
		linkcolor = purple,
		anchorcolor = purple,
		citecolor = purple,
		filecolor = purple,
		urlcolor = purple
	}
}

\newcommand{\approach}{HGFuzzer}

\begin{document}

\title{Directed Greybox Fuzzing via Large Language Model}

\author{Hanxiang Xu}
\affiliation{%
  \institution{Huazhong University of Science and Technology}
  \country{China}
}
\email{xuhx@hust.edu.cn}

\author{Yanjie Zhao}
\affiliation{%
  \institution{Huazhong University of Science and Technology}
  \country{China}
}
\email{yanjie_zhao@hust.edu.cn}

\author{Haoyu Wang}
\affiliation{%
  \institution{Huazhong University of Science and Technology}
  \country{China}
}
\email{haoyuwang@hust.edu.cn}

\begin{abstract}
Directed greybox fuzzing (DGF) focuses on efficiently reaching specific program locations or triggering particular behaviors, making it essential for tasks like vulnerability detection and crash reproduction. However, existing methods often suffer from path explosion and randomness in input mutation, leading to inefficiencies in exploring and exploiting target paths. In this paper, we propose \approach{}, an automatic framework that leverages the large language model (LLM) to address these challenges. 
\approach{} transforms path constraint problems into targeted code generation tasks, systematically generating test harnesses and reachable inputs to reduce unnecessary exploration paths significantly. Additionally, we implement custom mutators designed specifically for target functions, minimizing randomness and improving the precision of directed fuzzing.
We evaluated \approach{} on 20 real-world vulnerabilities, successfully triggering 17, including 11 within the first minute, achieving a speedup of at least 24.8× compared to state-of-the-art directed fuzzers. Furthermore, \approach{} discovered 9 previously unknown vulnerabilities, all of which were assigned CVE IDs, demonstrating the effectiveness of our approach in identifying real-world vulnerabilities.
\end{abstract}



\maketitle

\section{Introduction}
Fuzzing is a widely adopted and highly effective technique for identifying software vulnerabilities by generating and executing numerous test cases to detect abnormal program behaviors~\cite{CHEN2018118,8371326}. Over the years, fuzzing has evolved from simple random testing to more sophisticated and targeted strategies. Among these evolutionary paths, \textbf{greybox fuzzing} has emerged as a prominent approach that balances efficiency and scalability~\cite{8863940} by combining elements of both blackbox testing with minimal program visibility and whitebox testing with comprehensive program analysis.

Most greybox fuzzers are traditionally coverage-guided, aiming to maximize overall program exploration. As the field advanced, researchers developed \textbf{directed greybox fuzzing (DGF)}, which focuses on specific target locations, such as bug-prone regions, making it particularly useful for scenarios like patch testing and crash reproduction~\cite{aflgo,10.1145/3597926.3598067}. Despite its targeted approach, DGF faces two key challenges that significantly hinder its efficiency: \textbf{the complexity of exploration and the randomness of exploitation}. The exploration phase often encounters the issue of path explosion, where numerous infeasible paths are unnecessarily explored, consuming excessive resources. In the exploitation phase, the inherent randomness of input generation and mutation frequently results in irrelevant inputs being tested and fails to satisfy the precise constraints required to trigger vulnerabilities~\cite{10.1145/3512345}.

To improve the efficiency of directed fuzzing, various solutions have been proposed~\cite{Huang2024TitanE,10646712,298028}. One approach focuses on optimizing guidance toward target locations by analyzing program characteristics, such as identifying deviation basic blocks~\cite{beacon,Windranger} or satisfying execution path constraints~\cite{272157,PDGF}. Another approach aims to generate inputs with a higher likelihood of reaching target locations, for example, by predicting reachable inputs~\cite{FuzzGuard} or instrumenting only the relevant parts of the code~\cite{selectfuzz}. However, existing methods that leverage static analysis and symbolic execution to evaluate path reachability and constrain execution~\cite{aflgo,PDGF} introduce additional computational overhead and struggle with scalability. This inefficiency delays the discovery of vulnerabilities and limits testing effectiveness, \textbf{highlighting the need for more intelligent and targeted strategies in both exploration and exploitation phases}.

Recent advances in artificial intelligence, particularly large language models (LLMs), offer new opportunities to address these challenges in DGF. LLMs have demonstrated remarkable capabilities in various domains, including natural language understanding, code generation, and program analysis~\cite{ji2023benchmarkingexplaininglargelanguage}. Their ability to understand code semantics, infer program logic, and generate contextually relevant code makes them particularly promising for tackling the core difficulties in DGF. Several recent works have begun
to leverage LLMs to address key challenges in fuzzing~\cite{xu2024largelanguagemodelscyber,10.1145/3695988}, such as driver synthesis~\cite{10.1145/3658644.3670396, xu2024ckgfuzzerllmbasedfuzzdriver}, input generation~\cite{Meng2024LargeLM,shi2024harnessinglargelanguagemodels}, and bug detection~\cite{10298538}. However, \textbf{the application of LLMs specifically to the dual challenges of exploration complexity and exploitation randomness in DGF remains largely unexplored}.

Building on these insights, we propose a novel framework, i.e., \approach{}, that leverages the reasoning and code generation capabilities of LLMs to improve the efficiency of DGF. \approach{} tackles the complexity of exploration by transforming the path constraint problem into a code generation task. Specifically, it employs static analysis to identify potential call chains of the target function and guides LLM to analyze an available call chain, infer execution conditions, and generate target harness. The harness constrains execution paths to focus on relevant code regions, thereby reducing unnecessary exploration and computational overhead. For the exploitation phase, \approach{} reduces randomness by systematically guiding input generation and mutation. It utilizes LLM to generate reachable seeds that satisfy the execution constraints derived from the call chain analysis. Additionally, \approach{} constructs target-specific mutators by analyzing the characteristics of the target function and the associated vulnerability. These custom mutators are tailored to efficiently trigger the target vulnerability by prioritizing input mutations that align with the conditions required to exploit the vulnerability. By integrating these components, \approach{} provides an automatic approach to improve the efficiency of DGF.

We evaluate \approach{} across a benchmark of 20 real-world vulnerabilities, comparing it against state-of-the-art directed greybox fuzzers, including AFLGo~\cite{aflgo}, Beacon~\cite{beacon}, and SelectFuzz~\cite{selectfuzz}. \approach{} demonstrates superior performance by successfully triggering 17 out of 20 vulnerabilities, compared to 5 by AFLGo and SelectFuzz and 6 by Beacon, with 11 vulnerabilities triggered within the first minute of fuzzing. For successfully triggered known vulnerabilities, it achieves at least a 24.8× speedup to baselines. \approach{} also achieved a 27.5\% improvement in target function hit rate over the best-performing baseline, demonstrating its ability to effectively guide execution paths toward target functions. Additionally, \approach{} identified 9 previously unknown vulnerabilities in two open-source libraries, all of which were assigned CVE IDs.

In summary, this paper makes the following contributions:
\begin{itemize}
    \item We propose a novel framework, \approach{}, that integrates LLM to improve the efficiency of directed greybox fuzzing.
    \item We transform path constraint analysis into a code generation task, introducing LLM-guided reachable seed generation to reduce exploration complexity and a target-specific mutator to mitigate exploitation randomness during fuzzing.
    \item We implement \approach{} and demonstrate its superiority over state-of-the-art directed fuzzers on real-world benchmarks, successfully discovering 9 previously unknown vulnerabilities with CVE IDs.
\end{itemize}


\section{Background}
\subsection{Directed Greybox Fuzzing (DGF)}
DGF is a targeted software testing approach that focuses on efficiently reaching specific program locations or exposing certain program behaviors, such as vulnerable code regions or critical execution paths~\cite{Wang_2023}. Unlike traditional coverage-guided fuzzing, which maximizes coverage indiscriminately, DGF prioritizes inputs closer to pre-defined targets, saving resources and improving efficiency. First introduced by Böhme et al. with AFLGo~\cite{aflgo}, operating in two phases: \textbf{exploration} and \textbf{exploitation}. During exploration, the fuzzer aims to uncover as many paths as possible, favoring seeds that trigger new paths to maximize the potential of reaching targets. Once sufficient paths are uncovered, the exploitation phase focuses on inputs closer to the targets, assigning them more energy to generate mutations that fulfill the testing goals. Recent advancements, such as SemFuzz~\cite{SemFuzz}, ParmeSan~\cite{ParmeSan} and FuzzGuard~\cite{FuzzGuard}, have employed techniques like natural language processing, sanitizer-based annotations and machine learning to automate target identification, further expanding DGF’s applicability to scenarios like patch testing, crash reproduction, and detecting complex vulnerabilities.

\subsection{LLM for Fuzzing}
Recent research has explored the application of LLMs in fuzzing, demonstrating their potential to enhance input generation, driver synthesis, and bug detection processes~\cite{huang2024largelanguagemodelsbased, 10.1145/3663529.3663784}. Unlike traditional fuzzing methods, which often rely on randomized or rule-based approaches, LLMs utilize their generative capabilities to produce diverse and contextually valid inputs. Tools such as TitanFuzz~\cite{deng2023large} and FuzzGPT~\cite{deng2024large} leverage LLMs to improve seed mutation and input diversity, enabling more effective exploration of complex software behaviors. LLMs have also been applied to automate fuzz driver synthesis, a critical step in fuzzing workflows. For instance, Zhang et al.~\cite{zhang2023understanding} demonstrate the effectiveness of GPT-4 in generating fuzz drivers for library APIs, significantly reducing manual effort. Similarly, InputBlaster~\cite{liu2023testinglimitsunusualtext} and ChatAFL~\cite{meng2024large} enhance fuzzing for mobile apps and network protocols, achieving higher bug detection rates and improved input validity. 
Despite these advances, the potential of LLMs to specifically address the challenges of exploration complexity and exploitation randomness in DGF has yet to be thoroughly investigated.


\subsection{Challenges and Motivations}

\textbf{Complexity of exploration.}~Existing approaches in the exploration phase aim to uncover as many execution paths as possible, as new paths increase the likelihood of reaching the target~\cite{aflgo,Hawkeye,Windranger}. This is particularly necessary when the initial seeds are far from the target. Some methods also leverage lightweight static analysis to evaluate the reachability of execution paths to the target and use symbolic execution to constrain path exploration~\cite{beacon,selectfuzz,PDGF}. 
However, these methods face the challenge of path explosion caused by exploring numerous infeasible paths that cannot reach the given target in a library. Furthermore, existing methods fail to directly utilize the semantic results of static analysis. Instead, they convert these results into constraints for symbolic execution, which introduces additional computation overhead.

\begin{figure}[htbp]
    \centering
    \includegraphics[width=0.95\linewidth]{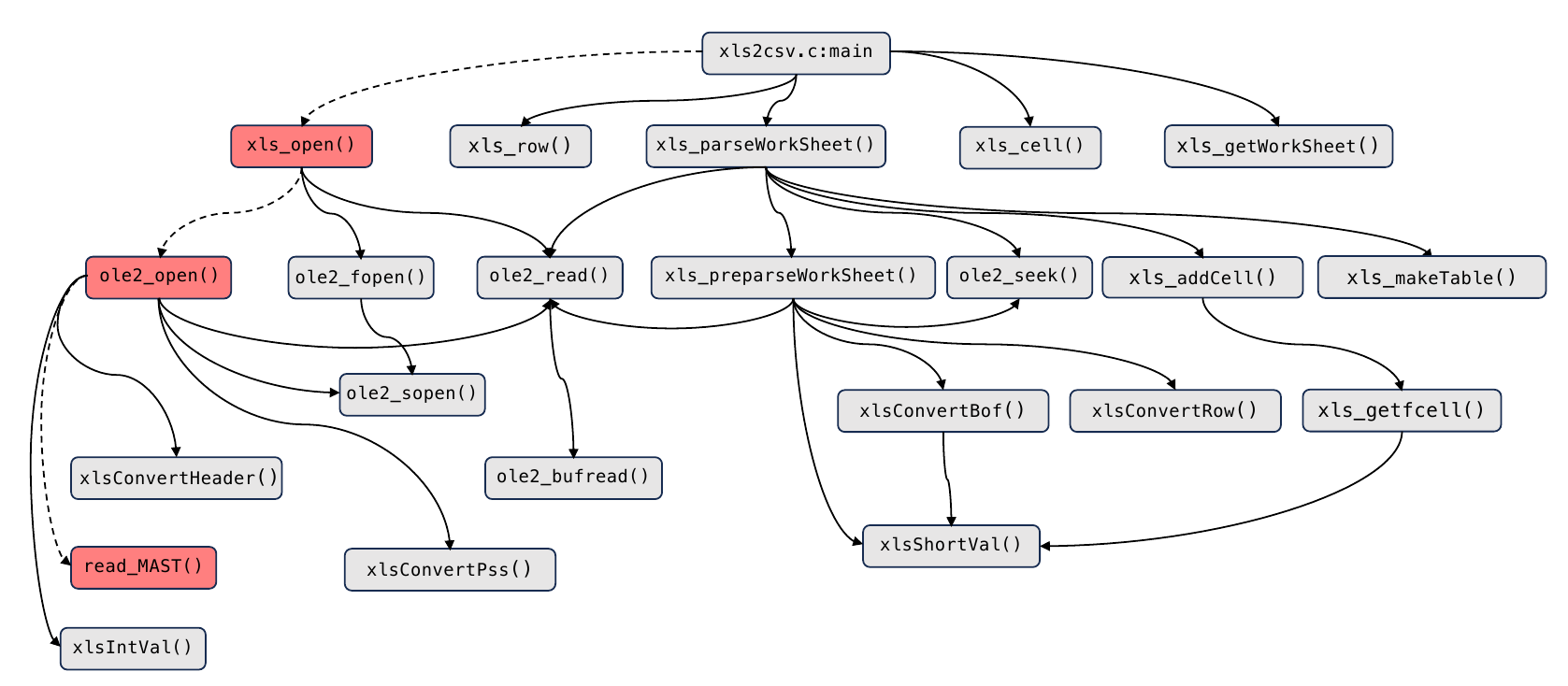}
    \caption{A motivating example from CVE-2017-2897, where the red blocks indicate a reachable path to the vulnerable target function \texttt{read\_MAST}.}
    \label{fig:motivation1}
\end{figure}

To address this issue, our core idea is to leverage the semantic understanding and code generation capabilities of LLM. We transform the complex path constraint problem into a code generation task. Specifically, \textbf{we collect the call chains of the target function within the target library, use LLM to analyze these call chains, and generate an executable harness that explicitly constrains the execution path to the target}. This approach avoids complex path exploration. As shown in~\autoref{fig:motivation1}, CVE-2017-2897 is an out-of-bounds write vulnerability in libxls 1.4.0. The entry program contains numerous complex execution paths. Traditional directed fuzzers require extensive path exploration before reaching the target, but most of these paths are irrelevant to the vulnerable function \texttt{read\_MAST}. By analyzing the call relations of \texttt{read\_MAST} in the library, we can generate a harness program to constrain the execution path explicitly, significantly simplifying the path exploration process.

\begin{figure}[htbp]
    \centering
    \includegraphics[width=0.95\linewidth]{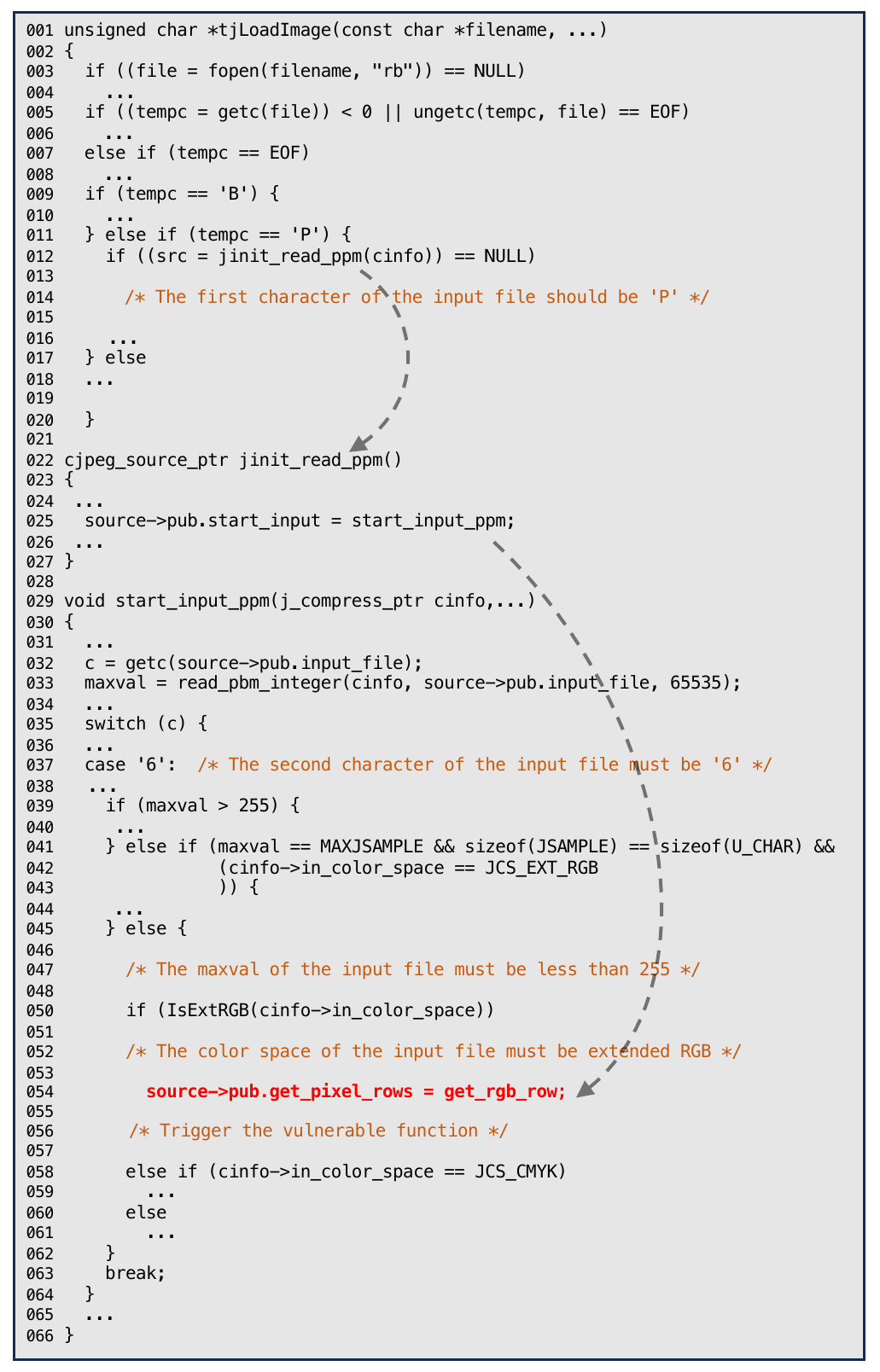}
    \caption{A motivating example from CVE-2020-13790.}
    \label{fig:motivation2}
\end{figure}

\noindent\textbf{Randomness of exploitation.}~Fuzzing is inherently a process driven by randomness, where the generation and mutation of inputs often rely on probabilistic techniques~\cite{10.1145/3512345,10.1145/3243734.3243804}. This randomness conflicts with the goal of directed fuzzing, as it can reduce efficiency by diverting resources to inputs that are less likely to reach or trigger the target during the exploitation phase.
As shown in~\autoref{fig:motivation2}, CVE-2020-13790 requires satisfying multiple path constraints to execute the vulnerable function \texttt{get\_rgb\_row}. For instance, the input file must start with specific characters (`P' followed by `6'), have a \texttt{maxval} less than 255, and use the \texttt{JCS\_EXT\_RGB} color space. Traditional directed fuzzers rely on input without confirmation of target reachability and random mutations to satisfy these constraints, which often results in numerous irrelevant inputs being tested, increasing the time required to reach or trigger the vulnerability. This randomness not only reduces efficiency but also leads to excessive resource consumption, as mutations are not explicitly guided toward satisfying the constraints necessary for exploitation. Addressing this challenge requires reducing randomness in the exploitation phase to better align with the goals of directed fuzzing.

To address this challenge, inspired by existing studies~\cite{Magneto,ISC4DGF,10.1145/3691620.3695513}, our basic idea is to utilize LLM to reduce the randomness in the exploitation process. Specifically, we first leverage LLM to analyze the complex parameter or variable constraints imposed by the target call chain, enabling the generation of initial reachable inputs for directed fuzzing. Subsequently, we use LLM to further analyze the characteristics of the target function, combined with the potential vulnerability risks, to generate custom mutators tailored for directed fuzzing. This approach aims to minimize the impact of randomness by systematically guiding the fuzzing process.

\section{Methodology}\label{sec:design}

\begin{figure*}[htbp]
    \centering
    \includegraphics[width=0.99\linewidth]{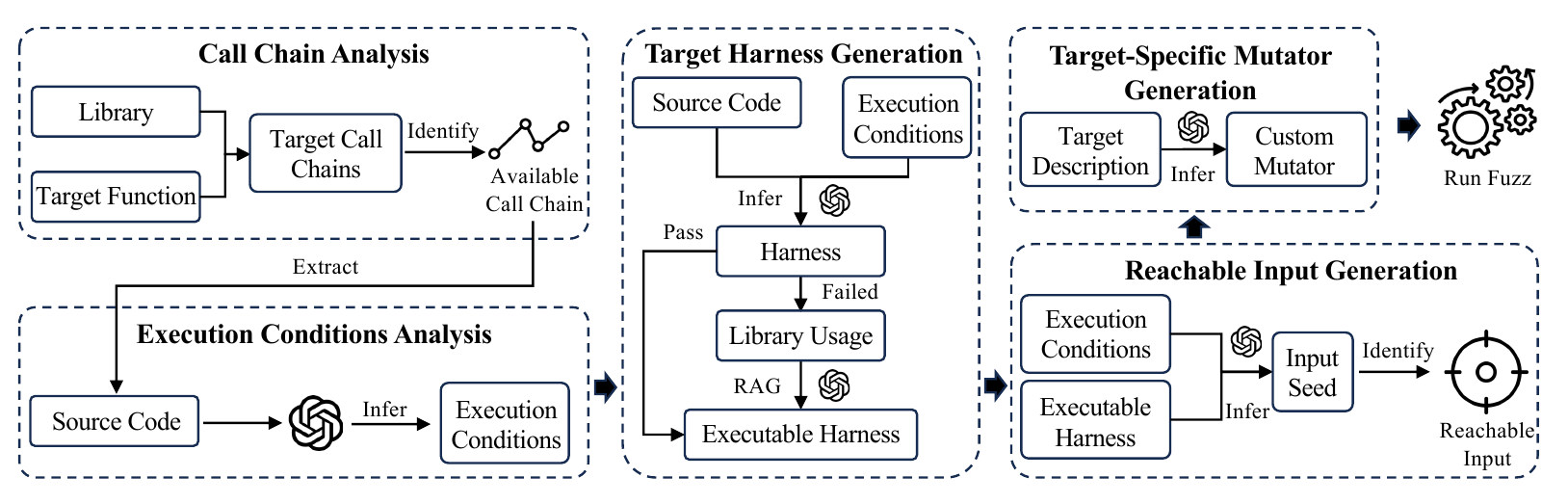}
    \caption{Overview of \approach\!.}
    \label{fig:workflow}
\end{figure*}

Our approach aims to improve the efficiency of directed fuzzing by leveraging the reasoning and code generation capabilities of LLM. As shown in~\autoref{fig:workflow}, \approach{} consists of multiple phases to systematically guide the fuzzing process toward specific target functions. First, it uses static analysis to identify all potential call chains of the target function and filters out the feasible ones~(\textit{\autoref{sec:3.1}}). Then, LLM analyze these call chains to infer the execution conditions required to trigger the target function~(\textit{\autoref{sec:3.2}}). Based on this information, \approach{} generates executable target harness and initial input seeds to guide execution paths toward the target~(\textit{\autoref{sec:3.3}},~\textit{\autoref{sec:3.4}}). Finally, \approach{} constructs target-specific mutators using target execution reports to optimize input mutation, reducing the time required to reach the target function and enhancing the effectiveness of the fuzzing process~(\textit{\autoref{sec:3.5}}).

\subsection{Call Chain Analysis}\label{sec:3.1}

Call chain analysis is the initial step of \approach{}, where we utilize static analysis to query all existing call chains of the target function within the library and extract an available call chain for further processing. This step provides crucial insights into the invocation context of the target function and forms the basis for subsequent reasoning and harness generation.

A call chain $C$ is represented as a sequence of functions $(F_n, F_{n-1}, \allowbreak \dots, \allowbreak F_1, F_0)$, where $F_0$ is the target function and $F_n$ is the starting function of the chain. Let $S$ denote the set of all call chains for the target function. To identify an available call chain, \approach{} parses $S$ following the criteria defined in~\autoref{alg:call_chain_identification}. If there exists a call chain $C \in S$ with $F_n = \texttt{main}$, \approach{} identifies the chain with the smallest length $|C|$ as the available call chain. If no such chain exists, \approach{} searches $S$ for the call chain with the smallest length $|C|$ whose starting function $F_n$ is declared as an external function in the library's header files. If such a chain is found, it is identified as the available call chain.

\begin{algorithm}[htp]
\caption{Call Chain Identification}
\label{alg:call_chain_identification}
\begin{algorithmic}[1]
\Require Target Function $F_0$, Library $L$
\Ensure Available Call Chain $C$
\State $S \gets \text{CallChains}(F_0, L)$
\State $C \gets \emptyset$, $l \gets \infty$ // $l$ means shortest length of available call chain
\For{$C' \in S$}
    \If{$F_n(C') = \texttt{main} \land |C'| < l$}
        \State $C \gets C'$, $l \gets |C'|$
    \EndIf
\EndFor
\If{$C = \emptyset$}
    \For{$C' \in S$}
        \If{$F_n(C') \in L.\text{headers} \land F_n(C') \text{ is extern} \land |C'| < l$}
            \State $C \gets C'$, $l \gets |C'|$
        \EndIf
    \EndFor
\EndIf
\State \Return $C$
\end{algorithmic}
\end{algorithm}

The preference for call chains starting with \texttt{main} is based on the observation that \texttt{main} and its associated file often provide complete parameter initialization, global variable declarations, and program constraints for calling entry function $F_{n-1}$. As the primary entry point, \texttt{main} is typically designed for specific functionalities, making it an ideal candidate template example for generating a target harness. 
When no call chain starting with \texttt{main} can be found, we prioritize call chains beginning with declared external functions, which typically have well-defined interfaces and parameter specifications. External functions, designed as entry points for libraries or modules, generally encapsulate the necessary contextual information and environment setup for correctly invoking target functionality. This strategy ensures we can identify call chains with sufficient context to build executable target harnesses even without \texttt{main}.
In summary, by prioritizing either \texttt{main}-originated chains or those starting with external functions, \approach{} ensures the selected call chain provides sufficient initialization context to generate valid and executable target harnesses.



\subsection{Execution Conditions Analysis}\label{sec:3.2}

Given an available call chain $C = (F_n, F_{n-1}, \dots, F_1, F_0)$, \approach{} then extracts the execution conditions required to traverse the chain and ultimately reach the target function $F_0$. This process involves parsing the source code of all functions in the call chain and using LLM to analyze the conditions for each function call in the chain.

\approach{} begins by utilizing static analysis to parse the source code of all functions in $C$. This parsing step generates a structured representation of the code, enabling precise identification of function definitions and call sites. For each pair of functions <$F_i, F_{i-1}$> in the chain, where $F_i$ calls $F_{i-1}$, \approach{} employs LLM to analyze the specific execution conditions required for the call. The LLM is designed to complete the following steps:

\begin{enumerate}
\item \textbf{Determining the Call Location}: Identifying the exact location in the source code where $F_{i-1}$ is called within $F_i$. This includes extracting the line number and the surrounding code snippet containing the call.
\item \textbf{Identifying Decision Variables}: Identifying the variables that determine whether the call from $F_i$ to $F_{i-1}$ occurs. These decision variables may include function parameters, global variables, or local variables within $F_i$.
\item \textbf{Analyzing Conditions}: Analyzing the conditions that these decision variables must satisfy for the call to occur. These conditions are represented as logical predicates or constraints on the variable values, such as inequalities, equality constraints, or ranges.
\end{enumerate}

\begin{figure}[htbp]
    \centering
    \includegraphics[width=0.85\linewidth]{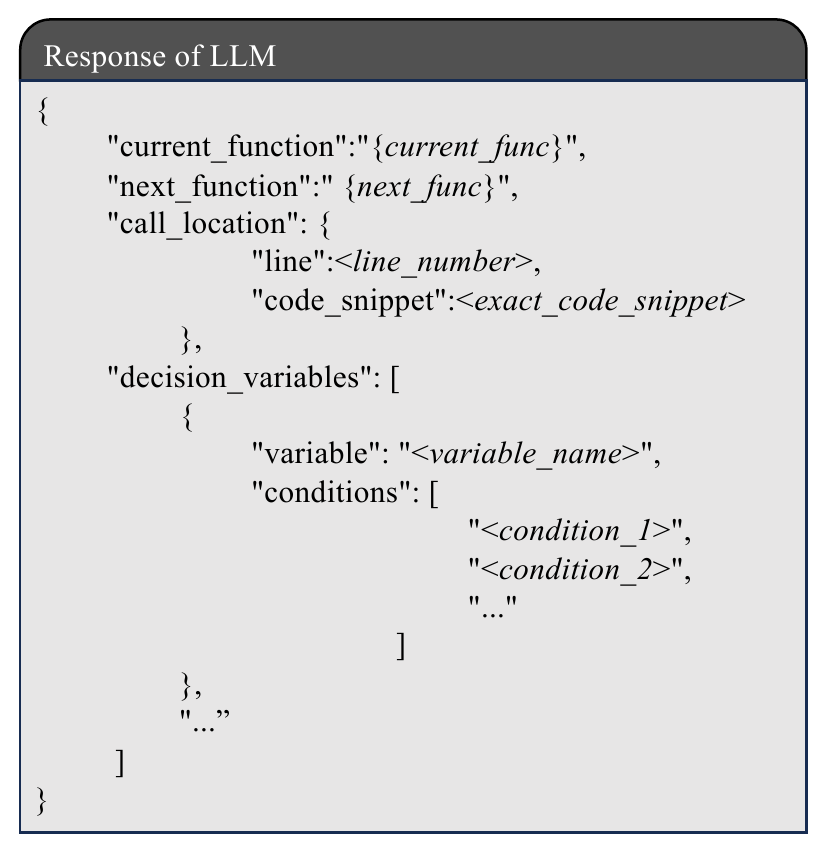}
    \caption{An example of the response from the LLM in execution conditions analysis.}
    \label{fig:output_format}
\end{figure}

\autoref{fig:output_format} illustrates the structured output of the LLM for a single function call <$F_i, F_{i-1}$>. Each decision variable is associated with its constraints, which are expressed in formal logic for clarity and precision. By iterating over all function pairs in the call chain, \approach{} builds a complete representation of the execution conditions needed to reach the target function $F_0$. This structured analysis accurately captures the dependencies and constraints of the call chain, providing essential information for generating executable harness and deriving inputs that can traverse the chain and trigger the target function.

\textbf{Illustrative Case.}~For the motivating example in~\autoref{fig:motivation2}, the call chain 
$C = (\texttt{tjLoadImage}, \texttt{jinit\_read\_ppm}, \texttt{start\_input\_ppm},$ \\
$\texttt{get\_rgb\_row})$ leads to the target function \texttt{get\_rgb\_row}. The call from 
\texttt{tjLoadImage} to \texttt{jinit\_read\_ppm} occurs at line 12 and is conditional on the first 
character of the input file ($\texttt{temp\_c}$) being $\texttt{`P'}$~(line 11). Within \texttt{start\_input\_ppm}, additional conditions must be satisfied, such as the second 
character of the file being $\texttt{`6'}$~(line 37), the maximum pixel value ($\texttt{maxval}$) being less 
than $255$~(line 39-45), and the color space of the input file being extended RGB ($\texttt{JCS\_EXT\_RGB}$)~(line 50). 
\approach{} identifies these decision variables and constraints for each function pair in the call 
chain, ensuring the conditions required to traverse the chain and execute the target function 
\texttt{get\_rgb\_row} are captured in a structured format for downstream use.

\subsection{Target Harness Generation}\label{sec:3.3}

To direct the fuzzing process toward the target function, \approach{} generates a targeted fuzz harness based on the available call chain, execution conditions, and corresponding function source code. This harness constrains the exploration space of the fuzzer, reducing redundant path exploration and ensuring efficient reachability of the target function.

\begin{figure}[htbp]
    \centering
    \includegraphics[width=0.9\linewidth]{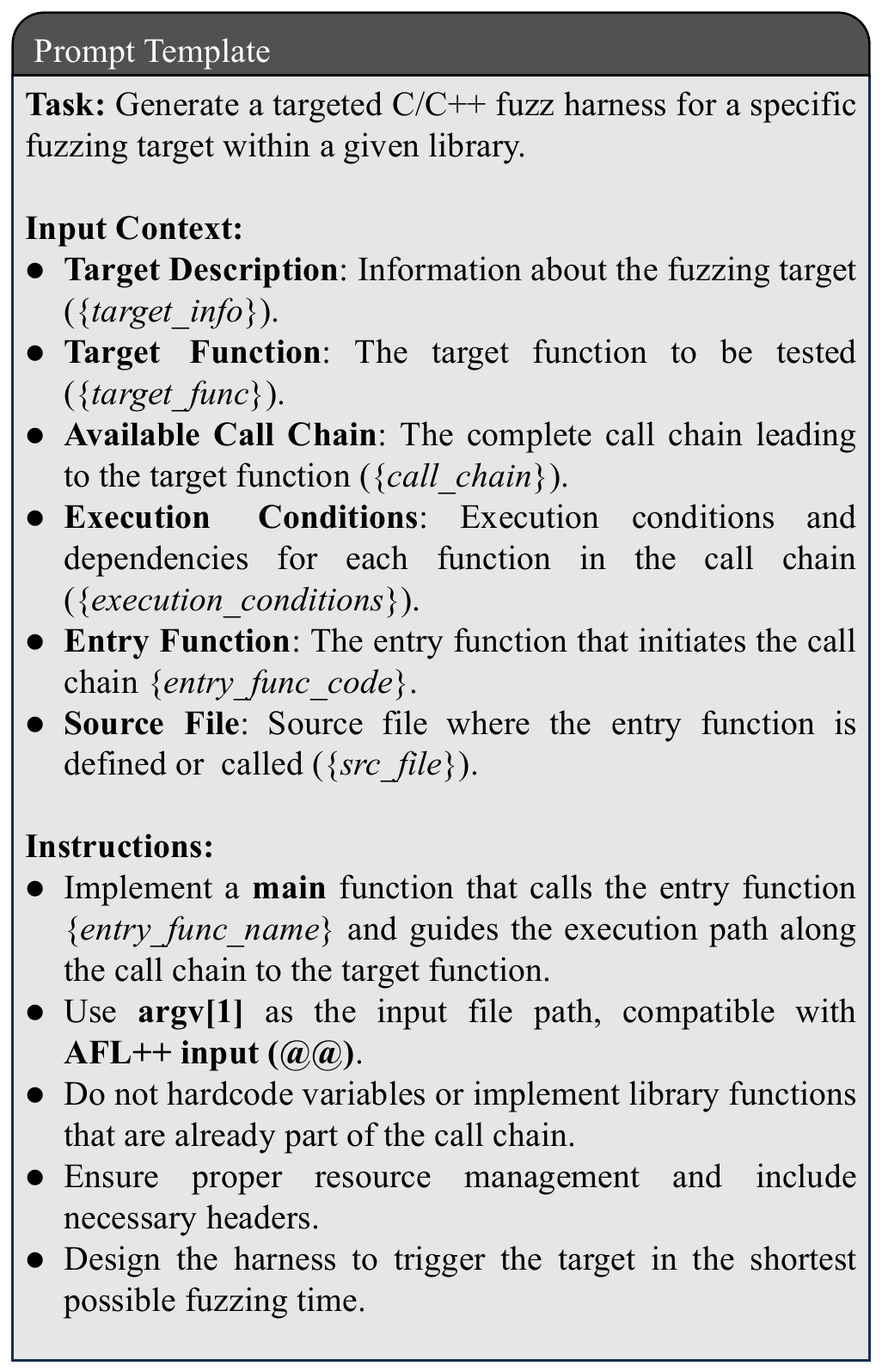}
    \caption{Prompt template for target harness generation.}
    \label{fig:harness}
\end{figure}

\autoref{fig:harness} presents the prompt template used in \approach{} for target harness generation. In addition to the available call chain, function source code, and execution conditions, the input context incorporates a target description, the entry function, and the corresponding source file. The target description provides an overall summary of the fuzzing objective, which may come from prior vulnerability reports (e.g., CVE records) or expert knowledge indicating potential weaknesses in the target function. This description helps guide the LLM in generating a relevant and effective harness. The entry function is determined based on the structure of the call chain. Given an available call chain $C = (F_n, F_{n-1}, \dots, F_1, F_0)$, if the final function $F_n$ is the \texttt{main} function, then the entry function is set to $F_{n-1}$, as it is the actual initiator of the call sequence. The \texttt{main} function is treated as a template for invoking $F_{n-1}$. In this case, the source file provided as input should be the file containing $F_n$ to ensure that the LLM has all necessary configuration details for calling $F_{n-1}$. If $F_n$ is not the \texttt{main} function, then $F_n$ itself is treated as the entry function, and the source file corresponds to the file where $F_n$ is defined. By structuring the prompt in this manner, \approach{} enables LLM to generate a valid fuzz harness that effectively triggers the target function.

\textbf{Compilation Error Resolution.}~To ensure that the generated fuzz harness can be successfully compiled into an executable program, \approach{} employs a compilation error resolution mechanism based on RAG (Retrieval-Augmented Generation), as shown in~\autoref{alg:rag_compilation_repair}. 
This mechanism is based on an external knowledge base that contains all source files, header files, and test cases from the target library. These files are embedded into vector representations and indexed to facilitate efficient similarity-based retrieval. If the harness fails to compile within the library's context, \approach{} collects the compilation errors issued by the compiler and constructs a query based on these errors. The query engine retrieves relevant code snippets from the knowledge base, including the usage, implementation, and definition of the error-related functions. The retrieved results are then used by the LLM to guide the repair process. Specifically, the LLM incorporates the context provided by the retrieved snippets to iteratively refine the harness until the compilation errors are resolved. This iterative repair process ensures that the harness is not only syntactically correct but also semantically compatible with the target library.

\begin{algorithm}[htbp]
\caption{Compilation Error Resolution with RAG.}
\label{alg:rag_compilation_repair}
\begin{algorithmic}[1]
\Require Compilation error $E$, Target Harness $H$, Knowledge base $K$, Embedding model $M$, Similarity threshold $s$, Number of chunks $k$, LLM $L$, Refinement prompt $P_r$
\Ensure Revised harness $H_{revised}$
\State $Q \gets BuildQuery(E)$ 
\State $IndexBase \gets Embed(K, M)$
\State $Q_{vec} \gets Embed(Q, M)$
\State $C_1, C_2, \dots, C_k \gets RetrieveChunks(Q_{vec}, IndexBase, s, k)$
\State $R \gets LLM(Q, C_1)$
\ForAll{remaining chunk $C_i$ in $C_2, \dots, C_k$}
    \State $R \gets RefineWithLLM(R, C_i, P_r, L)$
\EndFor
\State $H_{revised} \gets LLM(Q, R, H)$
\State \Return $H_{revised}$
\end{algorithmic}
\end{algorithm}

\subsection{Reachable Input Generation}\label{sec:3.4}

The target harness generated by \approach{} explicitly constrains the fuzzer's path exploration. However, the harness alone cannot directly guide the execution path to the target function. This limitation arises because the entry function often contains multiple branches, which create diverse execution paths. To ensure that the execution path reaches the target function, it is necessary to generate a reachable seed input that satisfies all execution conditions within the call chain. These execution conditions provide logical or mathematical constraints that must be met for the program to traverse the intended path. Given the diversity and complexity of library functionalities and their input formats, generating such an input manually is both labor-intensive and error-prone.

\begin{figure}[htbp]
    \centering
    \includegraphics[width=0.9\linewidth]{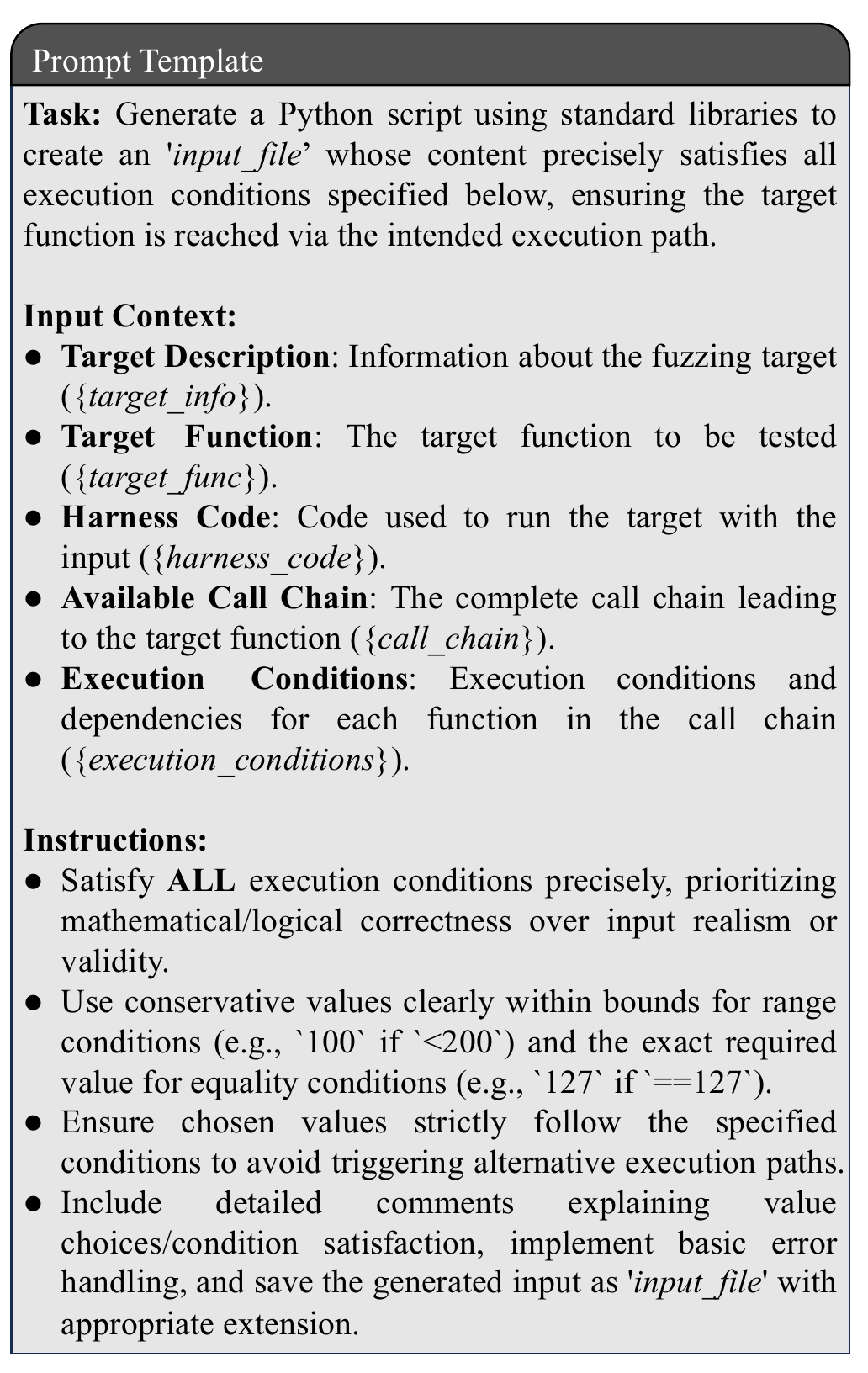}
    \caption{Prompt template for reachable input generation.}
    \label{fig:input}
\end{figure}

An alternative approach might be to employ LLMs to directly generate the initial input. However, this approach presents significant challenges when considering the complex input requirements of many libraries. For example, libraries such as \texttt{libming} process SWF or ABC flash files, \texttt{libjpeg-turbo} requires correctly formatted JPEG or PPM images, and other libraries may demand inputs with complex binary structures, specific headers, checksums, and format-specific requirements. Directly generating such specialized binary or structured data as raw content exceeds the current capabilities of LLMs and would invariably produce invalid inputs. For instance, creating a valid SWF file requires comprehensive knowledge of the flash file format specification, including its tag structure, binary encoding rules, and internal consistency requirements.

To address this challenge, \approach{} instead employs LLM to generate a Python script tailored to produce an initial input capable of meeting all specified execution conditions. This approach \textbf{leverages existing libraries in Python that can handle the complexities of file format specifications}, allowing for precise construction of valid inputs. To guide the LLM in generating the required Python script, \approach{} employs a structured prompt, as listed in~\autoref{fig:input}. The prompt supplies essential input context, encompassing the execution conditions derived from the program's call chain and the pertinent harness code. It explicitly defines the task for the LLM: to generate a Python script producing an input that satisfies all specified execution conditions, thereby ensuring the target function is reached while avoiding unintended execution paths. Additionally, the prompt details how to handle specific condition types, instructing the LLM to use values well within bounds for range constraints and precise values for equality constraints, leveraging Python libraries for accurate input construction.

Once the Python script is generated, it is executed to produce an initial input for the target harness. However, given the potential for the LLM to generate incorrect or suboptimal code~\cite{liu2023codegeneratedchatgptreally,pearce2021asleepkeyboardassessingsecurity,qiu2024efficientllmgeneratedcoderigorous}, \approach{} integrates a verification mechanism using afl-cov~\cite{afl-cov}. This tool instruments the target harness and monitors the execution path to verify whether the generated input successfully guides the program to the target function. If the input fails to reach the target function, the process is repeated iteratively. During each iteration, the prompt is refined, and the LLM generates a new script aiming to address previous deficiencies. This iterative process continues until a reachable input is obtained. By combining the generative capabilities of LLM with the verification of afl-cov, \approach{} ensures that the generated inputs effectively guide the execution path to the target function. 

\subsection{Target-Specific Mutator Generation}
\label{sec:3.5}

After generating a reachable input to guide the execution path to the target function, the next challenge is to ensure that the target function can be triggered. Triggering the target requires the program to enter specific execution states that satisfy the conditions necessary for exploiting the target vulnerability. To reduce the randomness of the fuzz engine in mutating the reachable seed, \approach{} leverages LLM to generate \textbf{a target-specific custom mutator}. The custom mutator is then integrated into the fuzzing engine to implement tailored mutation strategies designed for the specific target.

The generation of the custom mutator is based on a carefully designed prompt, which incorporates the target description, the target function's source code, and the reachable input script. Additionally, the prompt provides the LLM with the custom mutator API documentation and examples from the fuzzing engine to ensure compatibility. We guide the LLM through a three-step chain-of-thought prompt to generate the custom mutator:

\begin{enumerate}

    \item \textbf{Analyzing Root Cause}: The LLM first analyzes the target function's source code and the provided description of the target, same as mentioned in~\autoref{sec:3.3}. This analysis focuses on identifying the root cause of the vulnerability and the specific program states that must be reached to trigger it. By understanding the vulnerability pattern and its triggering conditions, the LLM gathers critical insights into how the input should be mutated to exploit the vulnerability.

    \item \textbf{Designing Mutation Strategy}: Based on the analysis of the target function and its triggering conditions, the LLM designs a mutation strategy aimed at efficiently exploiting the vulnerability. This strategy prioritizes mutations that directly target the vulnerability-triggering conditions. The approach ensures that input values are mutated toward extreme bounds or specific ranges that satisfy the conditions identified in the target function, while simultaneously avoiding mutations that would cause the input to fail the call chain conditions or deviate from paths leading to the target.

    \item \textbf{Generating Custom Mutator Code}: After defining the mutation strategy, the LLM generates the complete custom mutator code in C/C++, adhering to the custom mutator API. The mutator code implements the tailored mutation strategy specific to the target vulnerability while ensuring compatibility with the fuzzing engine. It includes all necessary headers, struct definitions, and initialization functions required by the custom mutator API documentation. The implementation maintains performance optimization to ensure the fuzzing engine runs smoothly without runtime errors or performance degradation.

\end{enumerate}



\begin{figure}[htbp]
    \centering
    \includegraphics[width=0.95\linewidth]{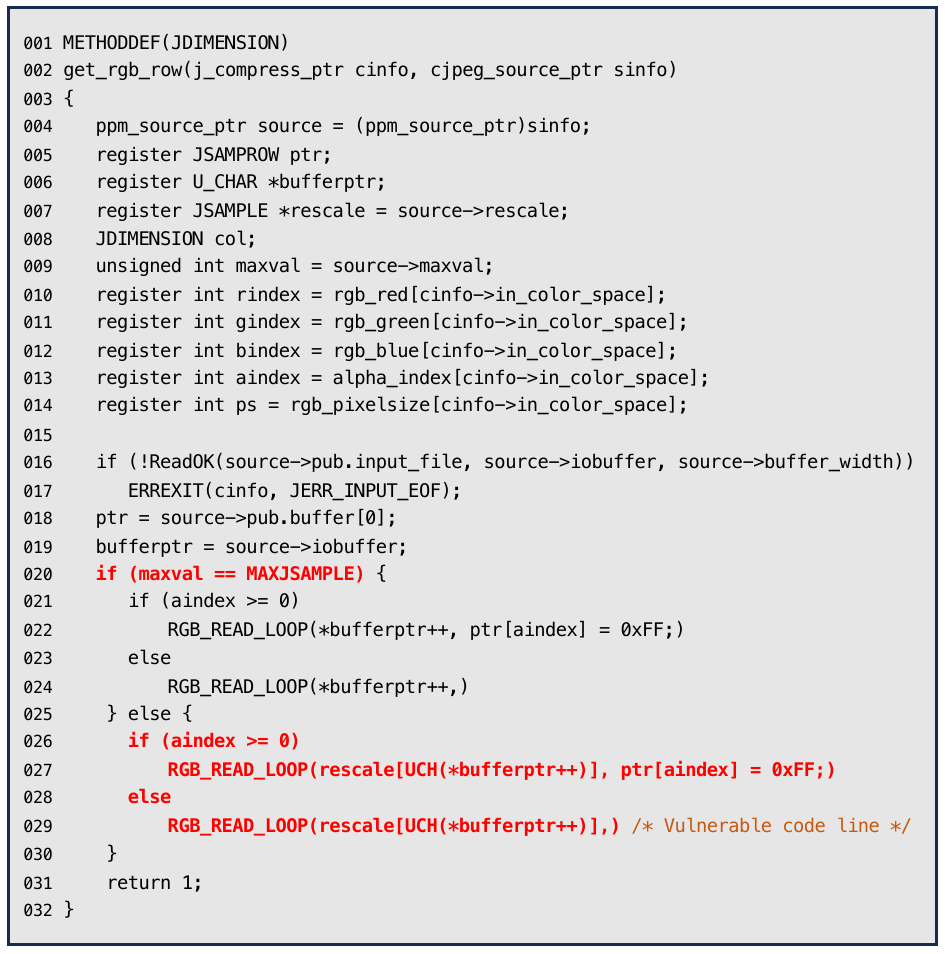}
    \caption{Vulnerable function in CVE-2020-13790.}
    \label{fig:mutator}
\end{figure}

To ensure the generated custom mutator is functional and effective, our approach incorporates an iterative refinement process. If the generated custom mutator fails to compile or causes runtime errors, the LLM automatically references the custom mutator API documentation to fix the issues. 
Additionally, our framework performs a lightweight validation check after integrating the custom mutator with the fuzzing engine, measuring basic execution metrics over a short sample run to verify that it operates without significant overhead or instability. 
This iterative process ensures that the custom mutator can efficiently guide the fuzzing process toward triggering the target vulnerability.


\textbf{Illustrative Case.}~To address the heap buffer overflow vulnerability in the \texttt{get\_rgb\_row} function from CVE-2020-13790 in~\autoref{fig:mutator}, the LLM first analyzes the source code and identifies that the root cause lies in the lack of proper boundary checks when accessing the \texttt{rescale} array through the pointer \texttt{bufferptr}. This occurs under specific conditions, such as \texttt{maxval!= MAXJSAMPLE}~(line 20-25) and \texttt{aindex<0}~(line 26-28), while processing malformed PPM files. Based on this analysis, the LLM designs a mutation strategy that includes generating malformed PPM headers, manipulating key variables (e.g., \texttt{maxval} and \texttt{aindex}), and targeting boundary conditions to trigger the vulnerability. The custom mutator generated by the LLM implements this strategy by mutating input files to include unexpected dimensions, non-standard pixel values, and insufficient data sizes. This allows the fuzzing engine to efficiently explore paths leading to the vulnerability, reducing randomness and improving the likelihood of triggering the overflow condition.

\subsection{Implementation}

We implemented \approach{} using approximately 3,400 lines of Python code and 500 lines of bash scripts. Our implementation is based on AFL++~\cite{afl++} due to its extensibility and robust feature set. For our LLM selection, we utilize Anthropic's API to access Claude-3.5-Sonnet~\cite{claude}. For static analysis, we used CodeQL~\cite{codeql} and Tree-Sitter~\cite{tree} to extract information about call chains and source code of functions. Additionally, we employed LlamaIndex~\cite{llamaindex} to build the RAG engine, enabling an efficient construction of prompts and retrieval of relevant program context.
\section{Evaluation}
In this section, we evaluate \approach{} using real-world vulnerabilities and aim to answer the following research questions:

\begin{itemize}
    \item \textbf{RQ1}: How effective is \approach{} in triggering known vulnerabilities?
    \item \textbf{RQ2}: Can \approach{} effectively constrain execution paths to hit the target functions more frequently?
    \item \textbf{RQ3}: How do the individual components of \approach{} contribute to its overall performance?
    \item \textbf{RQ4}: Can \approach{} detect new real-world vulnerabilities?
\end{itemize}

\noindent\textbf{Baselines.}~For evaluation, we compared \approach{} against three state-of-the-art greybox directed fuzzers: AFLGo~\cite{aflgo}, Beacon~\cite{beacon}, and SelectFuzz~\cite{selectfuzz}. 
\begin{itemize}
    \item \textbf{AFLGo}: A widely used directed fuzzer that guides the fuzzing process toward the target location by leveraging distance metrics to prioritize execution paths.
    \item \textbf{Beacon}: This fuzzer focuses exclusively on execution paths that can reach the target location, prioritizing exploration of relevant code paths.
    \item \textbf{SelectFuzz}: This fuzzer detects code regions related to the target location and focuses on exploring these regions to improve the likelihood of reaching the target.
\end{itemize}

Since \approach{} is built on AFL++, we also included AFL++, a coverage-guided fuzzer, as a baseline to evaluate the effectiveness of our enhancements. We also attempted to compare \approach{} with state-of-the-art directed fuzzers such as Titan~\cite{Huang2024TitanE}, PGDF~\cite{PDGF}, and DeepGo~\cite{deepgo}. However, Titan is designed for multi-target scenarios and encountered compatibility issues when applied to our single-target benchmark. Additionally, PGDF and DeepGo do not provide source code or documentation, making it infeasible for us to reproduce them.

\noindent\textbf{Benchmark Dataset.}~We constructed our benchmark dataset by collecting 20 CVE vulnerabilities sourced from prior fuzzing research work~\cite{aflgo,beacon,selectfuzz} and open-source libraries integrated with OSS-Fuzz~\cite{oss-fuzz}. These vulnerabilities span 12 versions of 9 open-source C/C++ libraries. As shown in~\autoref{tab:RQ1}, the dataset includes common vulnerability types in C/C++ programs and covers diverse functionalities of open-source libraries. These vulnerabilities were used to evaluate the ability of the baselines to perform directed fuzzing on specific targets. To ensure a fair comparison, we used the same compilation options and execution commands for all experiments. To mitigate the effects of fuzzing randomness on the results, each experiment was repeated 5 times, with a time budget of 24 hours per run.

\noindent\textbf{Environment.}~All experiments were conducted on a server equipped with an AMD 64-Core Processor and 1TB of RAM, running a 64-bit version of Ubuntu 22.04 LTS.

\begin{table*}[htbp]
\centering
\caption{Comparison of effectiveness with baselines on our benchmark. The ``Total'' row represents the number of vulnerabilities successfully triggered within the time budget.}
\renewcommand{\arraystretch}{1.15} 
\Huge
\resizebox{0.95\linewidth}{!}{%
\begin{tabular}{cllllccccccccc}
\toprule
\multirow{2}{*}{\textbf{Library}} & \multirow{2}{*}{\textbf{CVE ID}} & \multirow{2}{*}{\textbf{Vulnerability Location}} & \multirow{2}{*}{\textbf{Vulnerability Type}} & & \multicolumn{3}{c}{\textbf{\approach}} & \multicolumn{2}{c}{\textbf{AFLGo}} & \multicolumn{2}{c}{\textbf{Beacon}} & \multirow{2}{*}{\textbf{SelectFuzz}} & \multirow{2}{*}{\textbf{AFL++}} \\
\cmidrule(lr){6-8} \cmidrule(lr){9-10} \cmidrule(lr){11-12}
 & & & & & \textbf{TS} & \textbf{TTR} & \textbf{TTE} & \textbf{TS} & \textbf{TTE} & \textbf{TS} & \textbf{TTE} & & \\
\midrule
libming 0.4.7 & CVE-2016-9831 & util/parser.c:parseSWF\_RGBA & Buffer Overflow & & 6.96s & 104s & \textbf{I.E.} & 2,290s & 6.7h & 16s & 4.3h & 1.1h & 0.45h \\
\cmidrule{1-14}
\multirow{2}{*}{libming 0.4.8} & CVE-2017-9988 & util/parser.c:parseABC\_NS\_SET\_INFO & NULL Pointer Dereference & & 6.24s & 179s & \textbf{10.1h} & 3,265s & T.O. & 404s & T.O. & T.O. & T.O. \\
 & CVE-2018-13066 & util/parser.c:parseSWF\_DEFINETEXT & Memory Leak & & 6.28s & 94s & \textbf{I.E.} & 3,473s & T.O. & 131s & 5.8h & T.O. & T.O. \\
\cmidrule{1-14}
\multirow{3}{*}{libpng 1.6.34} & CVE-2018-13785 & pngrutil.c:png\_check\_chunk\_length & Integer Overflow & & 6.78s & 89s & \textbf{0.27h} & 563s & T.O. & 43s & T.O. & T.O. & T.O. \\
 & CVE-2018-14048 & png.c:png\_free\_data & Segmentation Violation & & 6.79s & 96s & T.O. & 573s & T.O. & 78s & T.O. & T.O. & T.O. \\
 & CVE-2019-7317 & pngerror.c:png\_safe\_execute & Use-after-free & & 7.12s & 107s & \textbf{I.E.} & 607s & T.O. & 98s & T.O. & T.O. & T.O. \\
\cmidrule{1-14}
libjpeg-turbo 2.0.1 & CVE-2018-20330 & turbojpeg.c:tjLoadImage & Buffer Overflow & & 6.2s & 71s & \textbf{2.17h} & U.A. & / & U.A. & / & U.A. & U.A. \\
\cmidrule{1-14}
libjpeg-turbo 2.0.4 & CVE-2020-13790 & rdppm.c:get\_rgb\_row & Buffer Overflow & & 6.25s & 77s & \textbf{3.12h} & 91s & T.O. & 22s & T.O. & T.O. & T.O. \\
\cmidrule{1-14}
libjpeg-turbo 2.0.9 & CVE-2021-46822 & rdppm.c:get\_word\_rgb\_row & Buffer Overflow & & 6.52s & 82s & T.O. & 697s & T.O. & 14s & T.O. & T.O. & T.O. \\
\cmidrule{1-14}
lcms2.9 & CVE-2018-16435 & src/cmscgats.c:AllocateDataSet & Integer Overflow & & 6.57s & 98s & \textbf{I.E.} & 950s & T.O. & 82s & T.O. & T.O. & T.O. \\
\cmidrule{1-14}
\multirow{4}{*}{libxls 1.4.0} & CVE-2017-2896 & src/xls.c:xls\_mergedCells & Out-of-bounds Write & & 6.33s & 102s & \textbf{I.E.} & 41s & I.E. & 11s & 0.35h & 3.7h & 0.85h \\
 & CVE-2017-2897 & src/ole.c:read\_MSAT & Out-of-bounds Write & & 6.31s & 99s & \textbf{I.E.} & 55s & 0.35h & 9s & I.E. & 0.25h & 0.23h \\
 & CVE-2017-2910 & src/xls.c:xls\_addCell & Out-of-bounds Write & & 6.09s & 104s & 0.13h & 64s & 1.2h & 15s & \textbf{I.E.} & 0.73h & 1.1h \\
 & CVE-2021-27836 & src/xls.c:xls\_getWorkSheet & Segmentation Violation & & 6.22s & 103s & \textbf{I.E.} & 112s & 0.45h & 16s & 0.5h & 2.43h & 1.23h \\
\cmidrule{1-14}
libzip 1.2.0 & CVE-2017-12858 & lib/zip\_dirent.c:\_zip\_dirent\_read & Use-after-free & & 6.12s & 82s & \textbf{1.1h} & 251s & T.O. & 14s & T.O. & T.O. & T.O. \\
\cmidrule{1-14}
\multirow{2}{*}{libgd 2.3.2} & CVE-2021-38115 & src/gd\_tga.c:read\_header\_tga & Out-of-bounds Read & & 7.36s & 79s & \textbf{I.E.} & U.A. & / & 23s & T.O. & U.A. & T.O. \\
 & CVE-2021-40812 & src/gd\_bmp.c:\_gdImageBmpCtx & Out-of-bounds Read & & 6.41s & 112s & \textbf{I.E.} & 543s & T.O. & 24s & T.O. & T.O. & T.O. \\
\cmidrule{1-14}
\multirow{2}{*}{cjson 1.7.16} & CVE-2023-50471 & cJSON.c:cJSON\_InsertItemInArray & Segmentation Violation & & 6.59s & 69s & \textbf{I.E.} & U.A. & / & U.A. & / & T.O. & T.O. \\
 & CVE-2023-50472 & cJSON.c:cJSON\_SetValuestring & Segmentation Violation & & 6.57s & 67s & \textbf{I.E.} & 11s & T.O. & U.A. & / & T.O. & T.O. \\
\cmidrule{1-14}
libmodbus 3.1.6 & CVE-2024-36844 & src/modbus.c:send\_msg & Use-after-free & & 6.09s & 98s & T.O. & 113s & T.O. & 10s & T.O. & T.O. & T.O. \\
\midrule
\textbf{Total} & & & & & \multicolumn{3}{c}{\textbf{17/20}} & \multicolumn{2}{c}{5/20} & \multicolumn{2}{c}{6/20} & 5/20 & 5/20 \\
\bottomrule
\end{tabular}%
} \\
\normalsize 
\footnotesize{\textbf{I.E.}: Imiediately Exploit (Trigger the vulnerability in one minute), \textbf{T.O.}: Time Out (>24 hours), \textbf{U.A.}: Unavailable (Static analysis or runtime error)}
\label{tab:RQ1}
\end{table*}

\subsection{RQ1:~Effectiveness in Triggering Known Vulnerabilities}\label{sec:rq1}

To answer RQ1, we evaluated the effectiveness of \approach{} against the baselines on the benchmark. We used \textbf{the time to exploit (TTE)} as the metric for evaluating effectiveness. Additionally, we compared \textbf{the time spent on static analysis (TS)} between \approach{}, AFLGo, and Beacon. Since SelectFuzz performs seed execution to detect code regions related to the target during the fuzzing process, and both SelectFuzz and AFL++ only perform compilation and instrumentation during the preparation phase, their TS was not recorded. We also evaluated the time for \approach{} to \textbf{guide the execution path to the target function (TTR)}. For evaluating the performance of the baselines on the benchmark, we collected the vulnerable library programs mentioned in the CVE reports or the fuzz drivers provided by the libraries as the fuzzing entry points. The initial inputs for the baselines were set to the test inputs or fuzz inputs provided by the libraries.

\begin{table*}[htbp]
\centering
\caption{Comparison of target function hit rates with baselines on our benchmark.}
\resizebox{0.8\linewidth}{!}{%
\renewcommand{\arraystretch}{1.1}
\begin{tabular}{cccccc}
\toprule
\textbf{CVE ID} & \textbf{\approach} & \textbf{AFLGo} & \textbf{Beacon} & \textbf{SelectFuzz} & \textbf{AFL++} \\
\midrule
CVE-2016-9831 & \textbf{79.16\%(38/48)} & 15.71\%(383/2,437) & 23.21\%(13/56) & 34.88\%(30/86) & 14.64\%(115/785) \\
CVE-2017-9988 & \textbf{53.76\%(256/476)} & 3.68\%(126/3,420) & 1.57\%(17/1,080) & 0\%(0/428) & 1.39\%(73/5,223) \\
CVE-2018-13066 & \textbf{81.82\%(9/11)} & 3.57\%(118/3,298) & 5.56\%(36/647) & 0.76\%(3/390) & 9.48\%(452/4,763) \\
CVE-2018-13785 & 94.73\%(18/19) & 94.10\%(319/339) & 90.09\%(100/111) & \textbf{97.63\%(124/127)} & 96.71\%(294/304) \\
CVE-2018-14048 & \textbf{99.63\%(270/271)} & 90.15\%(119/132) & 94.07\%(127/135) & 96\%(120/125) & 93.03\%(187/201) \\
CVE-2019-7317 & \textbf{100\%(3/3)} & \textbf{100.00\%(208/208)} & \textbf{100.00\%(126/126)} & \textbf{100.00\%(58/58)} & \textbf{100.00\%(199/199)} \\
CVE-2018-20330 & \textbf{100.00\%(320/320)} & U.A. & U.A. & U.A. & U.A. \\
CVE-2020-13790 & 7.94\%(29/365) & 1.00\%(21/2,094) & 8.07\%(39/483) & \textbf{14.10\%(22/156)} & 4.19\%(137/3,264) \\
CVE-2021-46822 & \textbf{9.91\%(36/363)} & 1.03\%(16/1,543) & 3.07\%(18/585) & 0\%(0/146) & 0.74\%(20/2,669) \\
CVE-2018-16435 & \textbf{22.47\%(20/89)} & 9.26\%(39/421) & 2.52\%(11/436) & 5.77\%(3/52) & 14.29\%(7/49) \\
CVE-2017-2896 & \textbf{92.96\%(66/73)} & 2.42\%(8/330) & 13.19\%(76/576) & 4.72\%(35/742) & 13.69\%(66/482) \\
CVE-2017-2897 & \textbf{95.83\%(69/72)} & 83.75\%(299/357) & 90.84\%(119/131) & 94.77\%(290/306) & 79.91\%(346/433) \\
CVE-2017-2910 & \textbf{56.25\%(36/64)} & 38.96\%(247/634) & 30.88\%(42/136) & 43.99\%(267/607) & 36.42\%(287/788) \\
CVE-2021-27836 & 73.53\%(50/68) & 69.43\%(602/867) & 73.45\%(498/678) & 46.87\%(1047/2,234) & \textbf{75.42\%(801/1,062)} \\
CVE-2017-12858 & 88.14\%(721/818) & 90.77\%(492/542) & 75.70\%(349/461) & \textbf{99.28\%(139/140)} & 95.65\%(22/23) \\
CVE-2021-38115 & \textbf{100.00\%(17/17)} & U.A. & \textbf{100.00\%(5/5)} & U.A. & \textbf{100.00\%(5/5)} \\
CVE-2021-40812 & \textbf{100.00\%(13/13)} & 44.10\%(202/458) & 37.70\%(204/541) & 68.60\%(59/86) & 42.04\%(206/490) \\
CVE-2023-50471 & \textbf{80\%(8/10)} & U.A. & U.A. & 0\%(0/114) & 0\%(0/37) \\
CVE-2023-50472 & \textbf{100.00\%(3/3)} & 0\%(0/612) & U.A. & 0\%(0/102) & 0\%(0/39) \\
CVE-2024-36844 & \textbf{98.75\%(79/80)} & 4.65\%(2/43) & 7.14\%(3/42) & 14.29\%(1/7) & 18.75\%(12/64) \\
\midrule
\textbf{AVG} & \textbf{64.75\%(2,061/3,183)} & 18.05\%(3,201/17,735) & 28.62\%(1,783/6,229) & 37.22\%(2,198/5,906) & 15.46\%(3,229/20,880) \\
\bottomrule
\end{tabular}%
}
\label{tab:RQ2}
\end{table*}

As detailed in~\autoref{tab:RQ1}, \textbf{\approach{} successfully triggered 17 out of 20 vulnerabilities within the 24-hour time budget}, the highest among all evaluated methods. In contrast, the baselines showed significantly lower success rates, with AFLGo and SelectFuzz each triggering 5 vulnerabilities and Beacon triggering 6, representing a 2.2x to 2.3x improvement in the number of successfully triggered vulnerabilities by \approach{}. The baselines also encountered cases where certain CVEs were marked as unavailable due to static analysis errors or runtime failures. For instance, AFLGo and Beacon failed to process CVE-2018-20330 due to block distance calculation errors, while SelectFuzz and AFL++ successfully initiated fuzzing on CVE-2018-20330 but encountered runtime errors, with their execution paths remaining constant at 1. This highlights the robustness of \approach{} in handling a wider range of vulnerabilities.

Among the vulnerabilities successfully triggered by \approach{}, \textbf{11 were exploited within the first minute of fuzzing}, showcasing its efficiency in reaching and triggering vulnerabilities. For example, in CVE-2016-9831, \approach{} exploited the vulnerability in under one minute, significantly faster than AFL++'s 0.45 hours, while the TTE for all other baselines exceeded 1 hour for this vulnerability. These results demonstrate that \approach{} achieves lower TTEs compared to the baselines, with differences often spanning orders of magnitude. This efficiency is primarily attributed to \approach{}'s ability to generate precise execution conditions and custom mutators tailored to the target function. By reducing randomness in input mutation and focusing on paths relevant to the vulnerability, \approach{} significantly accelerates the fuzzing process.

TS and TTR are influenced by factors such as project size, the complexity of the target function's call relationships, and the response time of the LLM API. However, experimental results show that \textbf{our approach has an advantage in processing time during the preparation phase}. Unlike AFLGo and Beacon, which incur significant computational overhead from calculating block distances during static analysis, \textbf{\approach{} only requires querying the call chains of the target function within the library}. This results in lower computational overhead and faster processing speed. For example, in \texttt{libming}, the average static analysis time for AFLGo exceeds 3000 seconds, while for Beacon, it exceeds 183 seconds. In contrast, our approach completes static analysis in just 6–7 seconds. Additionally, \approach{} guides the execution path to the target function within an average of \textbf{95.6 seconds}, avoiding the need for fuzz region guidance based on path distance calculations.

In conclusion, \approach{} effectively balances success rate, efficiency, and computational overhead in triggering known vulnerabilities. It outperforms baselines in both the number of vulnerabilities triggered and the time required to exploit them. Its ability to avoid costly block distance calculations during static analysis further highlights its efficiency, completing preparation phases significantly faster than baselines. These results emphasize \approach's practicality in triggering known vulnerabilities for open source libraries.

\subsection{RQ2:~Effectiveness in Constraining Execution Paths}

To evaluate whether \approach{} can effectively constrain execution paths to hit target functions more frequently, we examined how explicit path constraints generated by the LLM influence the fuzzing process. These path constraints are designed to reduce unnecessary exploration of non-target regions, potentially lowering performance overhead and improving the success rate of triggering vulnerabilities. For our analysis, we need to examine the execution paths explored during the experiments in \autoref{sec:rq1}.

During fuzzing, each unique execution path discovered generates a new seed that is preserved in the fuzzer's queue. Therefore, by analyzing these seeds, we can determine what proportion of the fuzzer's exploration successfully reached our target functions. We compared the target function hit rate across all approaches by using afl-cov to determine whether each generated seed hit the target function, then calculated the overall hit rate. This metric directly measures how effectively each approach guides execution toward the vulnerable code regions.


As illustrated in~\autoref{tab:RQ2}, \approach{} achieved the highest target function hit rate on 14 out of 20 CVEs. For example, in CVE-2016-9831, \approach{} achieved a hit rate of 79.16\%, significantly outperforming SelectFuzz (34.88\%) and AFL++ (14.64\%). Similarly, for CVE-2018-14048, \approach{} reached a hit rate of 99.63\%, which is higher than SelectFuzz (96\%) and other baselines. These results confirm the ability of \approach{} to guide execution paths effectively toward target functions. In contrast, the baselines demonstrated very low or even zero hit rates on certain CVEs, such as CVE-2018-13066 and CVE-2023-50472. This suggests that the library programs or their built-in fuzz drivers struggled or failed to reach the target functions, leading to poor performance of the baselines.

\approach{} also generated the smallest number of seeds across the benchmark dataset, with a total of 3,183 seeds. This indicates significantly fewer explored paths, demonstrating that \textbf{\approach{} effectively constrained the execution paths during fuzzing}. Compared to the best-performing baseline, SelectFuzz, which generated 5,906 seeds, \approach{} reduced the number of explored paths by 46\%. Despite exploring fewer paths, \approach{} achieved the highest average target function hit rate of 64.75\%, representing a 27.5\% improvement over SelectFuzz (37.22\%).

These results highlight three key findings: 
(1)~\approach{} significantly constrains the execution paths during directed fuzzing, reducing unnecessary exploration of non-target regions and improving efficiency. (2)~\approach{} achieves the highest target function hit rate, enabling it to fuzz the target functions more frequently, which correlates with its high success rate and low TTE observed in the~\autoref{sec:rq1} experiments. 
(3)~The use of LLM to explicitly generate target harnesses as fuzzing entry points demonstrates great flexibility, providing a significant advantage even in scenarios where open-source library programs struggle to reach the target functions. This flexibility ensures the effectiveness of \approach{} across a diverse range of programs and CVEs.

\subsection{RQ3:~Ablation Study}

\approach{} leverages the generative power of LLM to improve multiple key steps in existing directed fuzzing techniques. To evaluate the contributions of each component, we conducted an ablation study by defining three variations of \approach{}: (1) \approach{} without the reachable input generation component, using the same initial inputs as the baselines in~\autoref{sec:rq1}~(\textbf{Without Input}); (2) \approach{} without the target-specific mutator generation~(\textbf{Without Mutator}); and (3) \approach{} without both reachable input generation and target-specific mutator generation, relying solely on the target harness generation~(\textbf{Harness-only}). We evaluated the TTE of these variations across the benchmark to measure their effectiveness.

\begin{table}[htbp]
\centering
\caption{Comparison of effectiveness with different \approach{} variations.}
\resizebox{0.99\linewidth}{!}{%
\renewcommand{\arraystretch}{1}
\begin{tabular}{cccc}
\toprule
\textbf{CVE ID} & \textbf{Without Input} & \textbf{Without Mutator} & \textbf{Harness-only} \\
\midrule
CVE-2016-9831   & 0.28h                            & 0.1h                            & 0.25h                        \\
CVE-2017-9988   & T.O.                             & T.O.                            & T.O.                         \\
CVE-2018-13066  & T.O.                             & T.O.                            & T.O.                         \\
CVE-2018-13785  & 9.8h                             & 10.65h                          & 13.7h                        \\
CVE-2018-14048  & T.O.                             & T.O.                            & T.O.                         \\
CVE-2019-7317   & T.O.                             & T.O.                            & T.O.                         \\
CVE-2018-20330  & 5.32h                            & 2.3h                            & T.O.                         \\
CVE-2020-13790  & 7.65h                            & 9.1h                            & 17.5h                        \\
CVE-2021-46822  & T.O.                             & T.O.                            & T.O.                         \\
CVE-2018-16435  & T.O.                             & 0.12h                           & T.O.                         \\
CVE-2017-2896   & I.E.                             & I.E.                            & I.E.                         \\
CVE-2017-2897   & I.E.                             & 0.16h                           & 0.22h                        \\
CVE-2017-2910   & I.E.                             & 0.11h                           & 0.15h                        \\
CVE-2021-27836  & I.E.                    & 0.27h                           & 0.25h                        \\
CVE-2017-12858  & 7.5h                             & 2.45h                           & 7.9h                         \\
CVE-2021-38115  & I.E.                    & 0.13h                           & 0.05h                        \\
CVE-2021-40812  & I.E.                             & 0.08h                           & 0.1h                         \\
CVE-2023-50471  & I.E.                             & I.E.                            & I.E.                         \\
CVE-2023-50472  & I.E.                             & I.E.                            & I.E.                         \\
CVE-2024-36844  & T.O.                             & T.O.                            & T.O.                         \\
\midrule
\textbf{Total}  & 13/20                            & 14/20                           & 12/20                        \\
\bottomrule
\end{tabular}%
}
\label{tab:RQ3}
\end{table}

As shown in~\autoref{tab:RQ3}, the removal of individual components led to significant degradation in performance. The Without Input variation successfully triggered vulnerabilities in only 13 out of 20 CVEs, with longer TTE compared to the full \approach{} (e.g., 5.32h for CVE-2018-20330 compared to 2.17h in~\autoref{sec:rq1}). The Without Mutator variation performed slightly better, exposing vulnerabilities in 14 out of 20 CVEs, but its TTE was also noticeably higher (e.g., 10.65h for CVE-2018-13785 compared to 0.27h in~\autoref{sec:rq1}). The Harness-only variation showed the weakest performance, successfully exposing vulnerabilities in only 12 out of 20 CVEs, and often failed to reduce TTE to a practical level (e.g., timing out on CVE-2018-20330).

By comparing the results in~\autoref{tab:RQ1} and~\autoref{tab:RQ3}, the ablation experiments demonstrate the essential contributions of each component in \approach. The absence of reachable input generation significantly reduces the ability to guide execution toward the target function, as it ensures the satisfaction of execution conditions derived from the call chain analysis. Similarly, removing the target-specific mutator generation hinders efficient mutation of inputs to satisfy vulnerability-triggering conditions; the fuzzer relies on random mutations, leading to excessive exploration of irrelevant paths and increased TTE (e.g., 5.32h vs. 2.17h in CVE-2018-20330). The Harness-only variation, which lacks both components, performs the worst, as the harness alone cannot ensure correct path traversal or condition satisfaction, leading to lower success rates and frequent timeouts. We observe that for CVE-2023-50471 and CVE-2023-50472, the three variations exhibit no significant differences compared to the results in~\autoref{sec:rq1}, as their exploitability heavily depends on specific library execution paths. \approach{} effectively captures these features and incorporates them into the target harness, achieving good performance even with the Harness-only variation. These results highlight that \textbf{reachable input generation} and \textbf{target-specific mutators} are essential to reducing randomness and improving directed fuzzing efficiency, with the full integration of all components providing the best performance.

\subsection{RQ4:~Vulnerability Detection}
In this section, we apply \approach{} to detect new vulnerabilities. We selected two open-source libraries from the benchmark dataset (\texttt{libming} and \texttt{lcms}) and tested their latest versions after compilation. Since directed fuzzers require specific fuzzing targets, we adopted two strategies to define these targets. First, we investigated the most recent vulnerabilities in these libraries from the CVE database~\cite{cve} and used their root cause locations as fuzzing targets, which is based on the observation that certain vulnerable locations may contain multiple related bugs~\cite{Wang2020NotAC}. Second, we invited a security expert to manually review the code of the target libraries to identify additional suspicious locations. We applied \approach{} to each target for 24 hours of fuzzing. As a result, \approach{} discovered 9 new vulnerabilities across the two open-source libraries. All 9 vulnerabilities were assigned CVE IDs, demonstrating the real-world impact of our approach. The details of these findings are summarized in~\autoref{tab:RQ4}.

\begin{table}[htbp]
\centering
\caption{Vulnerabilities identified by \approach\!.}
\resizebox{\linewidth}{!}{%
\renewcommand{\arraystretch}{1.3} 
\Huge
\begin{tabular}{clcc}
\toprule
\textbf{Library} & \textbf{Vulnerability Location} & \textbf{Vulnerability Type} & \textbf{CVE} \\
\midrule
\multirow{8}{*}{libming 0.4.8} & util/parser.c:parseSWF\_EXPORTASSETS & Memory Leak & CVE-2025-26304 \\
 & util/parser.c:parseSWF\_SOUNDINFO & Memory Leak & CVE-2025-26305 \\
 & util/read.c:readSizedString & Memory Leak & CVE-2025-26306 \\
 & util/parser.c:parseSWF\_IMPORTASSETS2 & Memory Leak & CVE-2025-26307 \\
 & util/parser.c:parseSWF\_FILTERLIST & Memory Leak & CVE-2025-26308 \\
 & util/parser.c:parseSWF\_DEFINESCENEANDFRAMEDATA & Memory Leak & CVE-2025-26309 \\
 & util/parser.c:parseABC\_FILE & Memory Leak & CVE-2025-26310 \\
 & util/parser.c:parseSWF\_CLIPACTIONS & Memory Leak & CVE-2025-26311 \\
\midrule
lcms2.16 & cmsgamma.c:smooth2 & Buffer Overflow & CVE-2025-29070 \\
\bottomrule
\end{tabular}%
}
\label{tab:RQ4}
\end{table}

\begin{figure}[htbp]
\centering
\begin{subfigure}{\linewidth}
\includegraphics[width=\linewidth]{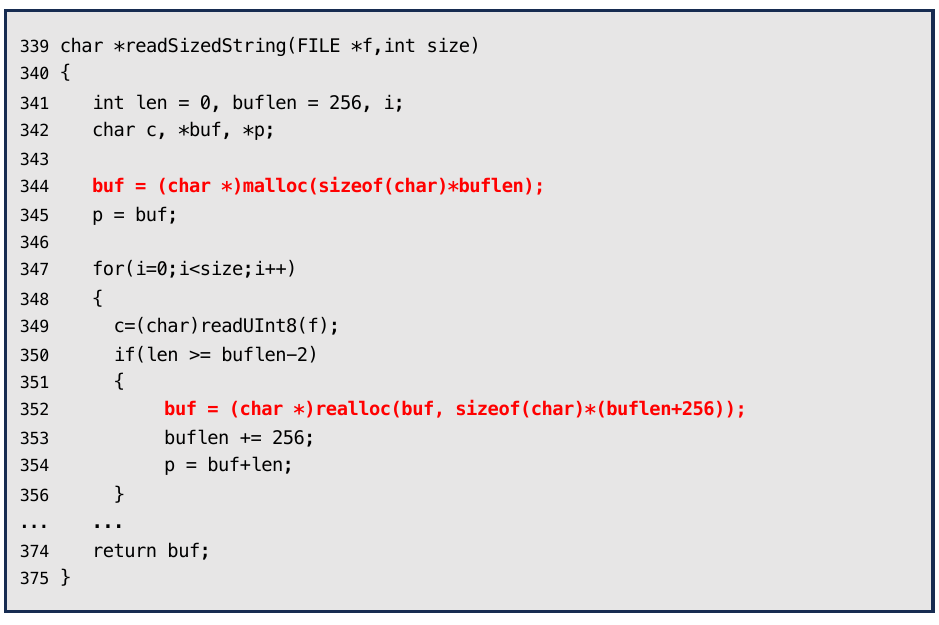}
\centering
\caption{A memory leak case~(CVE-2025-26306) in libming 0.4.8.}
\label{fig:case1}
\end{subfigure}
\begin{subfigure}{\linewidth}
\includegraphics[width=\linewidth]{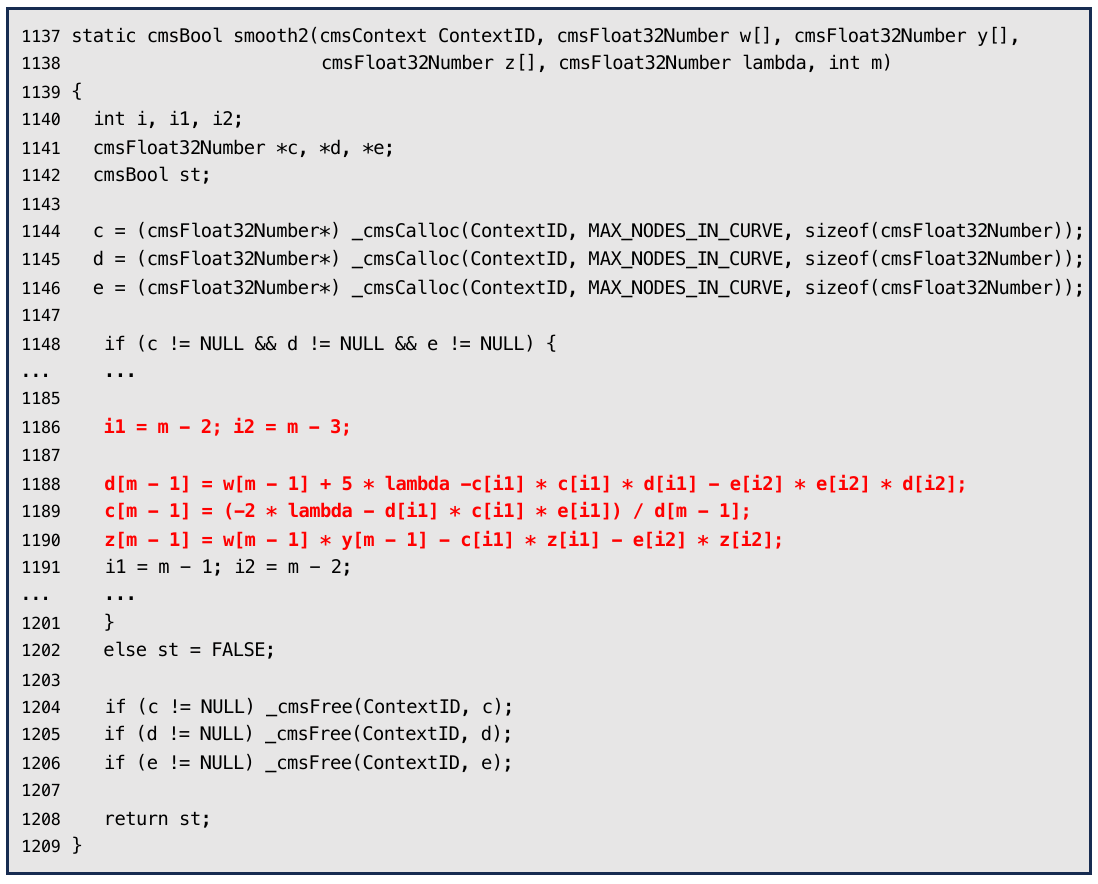}
\centering
\caption{A buffer overflow case~(CVE-2025-29070) in lcms2.16. }
\label{fig:case2}
\end{subfigure}
\caption{Examples of identified vulnerabilities.}
\label{fig:case}
\end{figure}

To further demonstrate the effectiveness to detect unknown vulnerabilities of \approach{}, we provide two representative examples of the vulnerabilities it identified, as shown in~\autoref{fig:case}. First, \approach{} identified a memory leak vulnerability in the \texttt{readSizedStr\\ing} function of libming 0.4.8. The issue arises from insufficient error handling during dynamic memory allocation. The function allocates a buffer using \texttt{malloc} (line 344) and resizes it with \texttt{realloc} (line 352) when the buffer size is exceeded. However, in cases where \texttt{realloc} fails, the previously allocated memory is not freed, leading to a memory leak. Additionally, the function lacks proper cleanup logic in error conditions, such as when \texttt{readUInt8} fails during the loop (lines 347–356). This vulnerability was assigned CVE-2025-26306. Second, \approach{} revealed a buffer overflow vulnerability in the \texttt{smooth2} function of lcms2. The function dynamically allocates memory for the \texttt{c}, \texttt{d}, and \texttt{e} arrays using \texttt{\_cmsCalloc} (lines 1144–1146) and subsequently accesses elements based on the input parameter \texttt{m}. At lines 1188–1190, the function performs calculations that access \texttt{d[m-1]}, \texttt{d[m-2]}, and \texttt{d[m-3]} without validating whether \texttt{m} is large enough to ensure safe access. If the input \texttt{m} is less than 4, these array accesses result in a buffer overflow. 
\section{Discussion}
\approach{} demonstrates the effectiveness of applying LLM to DGF. To explore more effective DGF approaches, we discuss core aspects of our approach and identify promising future directions.

\textbf{Selection of LLM.}~In our implementation of \approach{}, we utilized Claude-3.5-Sonnet as the underlying LLM. However, the performance of different LLMs may vary significantly, leading to discrepancies in outputs and results. More advanced LLMs generally exhibit a deeper understanding of the instructions provided in prompts and are capable of generating higher-quality harnesses and initial inputs. This suggests that the performance of \approach{} could be further enhanced by employing more powerful LLMs. Additionally, we observed during our experiments that hallucinations in the LLM could produce incorrect all-chain analysis results~\cite{10.1145/3703155}. This issue can lead to the generation of reachable inputs that fail to satisfy specific constraints. For example, in the case of complex conditional branches, as illustrated in~\autoref{fig:motivation2}, the LLM might mistakenly interpret a condition requiring \texttt{maxval} to be less than 255 as requiring \texttt{maxval} to be greater than or equal to 255. Such misinterpretations negatively impact the effectiveness of \approach{}, especially in scenarios with intricate logic or constraints.
A validation mechanism that cross-checks LLM-generated constraint interpretations against the original code could be implemented to address this issue, which we leave as a direction for future work.

\textbf{Expanding Applicability.}~The benchmark libraries used in our evaluation are open-source, with their source code publicly available. These libraries are also likely included in the training data of the selected LLM. As a result, applying \approach{} directly to closed-source libraries might yield different evaluation outcomes. Inspired by AFGen's approach of generating harnesses for whole functions of applications~\cite{10646877}, we also consider leveraging internal function call relationships as a reference for generating target harnesses. This could enable \approach{} to be applied in a broader range of software scenarios, such as Linux-based systems and web frameworks. Expanding the scope of \approach{} to these domains would require addressing challenges related to analyzing closed-source code and handling complex system-level dependencies, which presents an interesting direction for future work.

\section{Related Work}
\subsection{Directed Greybox Fuzzing}
DGF has recently gained attention due to its ability to efficiently locate specific target sites or trigger certain program behaviors. Böhme et al.~\cite{aflgo} proposed AFLGo, a pioneering DGF tool that utilizes a distance-based fitness metric to prioritize seeds closer to target locations. AFLGo calculates target distances during compile time and uses this information at runtime to guide the fuzzing process, achieving significant efficiency improvements in tasks such as bug reproduction. Building on AFLGo, tools like Hawkeye~\cite{Hawkeye} introduced function-level trace similarity to enhance seed prioritization, while tools such as UAFL~\cite{10.1145/3377811.3380386} and UAFuzz~\cite{259749} focused on detecting complex behavioral bugs like use-after-free vulnerabilities using sequence-aware fitness metrics. Berry~\cite{liang2020sequence} extended this concept by incorporating execution context into target sequences, improving the accuracy of target coverage.


To improve target identification, recent tools have integrated advanced techniques. SUZZER~\cite{2020Suzzer}, V-Fuzz~\cite{9199813}, and DeepG\-o~\cite{deepgo} employ deep learning models to predict vulnerable code regions, enabling the automatic labeling of potential targets without manual effort. AFLChurn~\cite{zhu2021regression} and DeltaFuzz~\cite{zhang2022deltafuzz} leverage code changes from version control systems to identify patch-related targets, making them suitable for regression testing scenarios. Additionally, tools like FuzzGuard~\cite{FuzzGuard} and BEACON~\cite{beacon} use lightweight static analysis or deep learning to filter out unreachable inputs, significantly improving fuzzing efficiency. For example, FuzzGuard has been shown to reduce unnecessary path executions by over 80\%, while BEACON combines symbolic execution with filtering mechanisms to further optimize performance. These advancements highlight the diverse approaches in DGF, focusing on improved target identification, advanced fitness metrics, and optimization strategies to enhance fuzzing performance across a variety of scenarios.

\subsection{LLM for Fuzzing}
Recent advancements in integrating LLM with fuzzing have demonstrated their potential to address challenges in generating meaningful and context-aware test cases for complex software systems~\cite{huang2024largelanguagemodelsbased,xu2024largelanguagemodelscyber,10.1145/3663529.3663784}. Unlike traditional fuzzing methods, which often rely on simple input generation techniques, LLM enable more sophisticated approaches by leveraging their generative capabilities. Tools such as TitanFuzz~\cite{deng2023large}, FuzzGPT~\cite{deng2024large}, WhiteFox~\cite{yang2023whitefox}, and ParaFuzz~\cite{yan2024parafuzz} utilize models like GPT~\cite{achiam2023gpt} and Codex~\cite{finnie2022robots} to improve input diversity and quality. These approaches incorporate LLM into prompt engineering and seed mutation processes, enhancing the effectiveness of fuzzing. Furthermore, LLMs have been applied to refine mutation strategies, beyond conventional techniques such as bit-flipping~\cite{wang2019superion}. E.g., CovRL-Fuzz~\cite{10.1145/3650212.3680389} employs LLM to perform coverage-guided mutations while maintaining input validity, increasing the probability of uncovering unexpected software behavior.

In addition to improving fuzzing inputs, LLM are increasingly employed to automate and simplify fuzzing workflows, particularly in the generation of fuzz drivers. InputBlaster~\cite{liu2023testinglimitsunusualtext} demonstrates the use of LLM to create specialized text inputs for mobile applications, achieving higher bug detection rates compared to conventional methods. Similarly, ChatAFL~\cite{meng2024large} and ChatFuzz~\cite{hu2023augmenting} integrate LLM to enhance fuzzing for web applications and network protocols, resulting in greater code coverage and better vulnerability identification. LLM have also proven effective in automating fuzz driver generation, which is critical for targeting specific APIs or software functions. Zhang et al.~\cite{zhang2023understanding} employ GPT-3.5 and GPT-4 to automate the creation of fuzz drivers for complex library APIs, reducing manual effort and achieving over 60\% automation. Similarly, CKGFuzzer~\cite{xu2024ckgfuzzerllmbasedfuzzdriver} leverages LLM along with a code knowledge graph to iteratively generate high-quality fuzz drivers, enabling deeper exploration of previously untested library code and improving code coverage. These advancements demonstrate the potential of integrating LLM with knowledge-driven frameworks to streamline fuzzing processes and enhance their overall effectiveness.
\section{Conclusion}
In this work, we present \approach{}, a novel and automatic framework that integrates LLM to enhance DGF. By transforming path constraint analysis into code generation tasks, \approach{} systematically generates target harnesses and reachable inputs, effectively reducing unnecessary path exploration. Additionally, it employs custom mutators tailored to specific vulnerabilities, minimizing randomness in input mutation and improving fuzzing precision. Evaluated on 20 real-world vulnerabilities, \approach{} outperformed state-of-the-art fuzzers, successfully triggering 17 vulnerabilities, 11 of which were triggered within the first minute, and achieving a speedup of at least 24.8× compared to baselines. Moreover, \approach{} discovered 9 previously unknown vulnerabilities, all of which received CVE IDs. These results demonstrate the effectiveness of \approach{} in improving fuzzing efficiency and precision.

\bibliographystyle{ACM-Reference-Format}
\bibliography{ref}

\end{document}